\documentclass[runningheads,hidelinks,pdftex]{llncs}
\usepackage{graphicx,url,hyperref,todonotes}
\usepackage{amsmath,amsfonts,amssymb,amstext,latexsym}
\usepackage{algorithm,algorithmicx}
\usepackage[noend]{algpseudocode}
\usepackage{listings}
\usepackage[pdftex]{color}
\usepackage{array}
\usepackage{microtype}

\algdef{SE}[SUBALG]{bindent}{eindent}{}{\algorithmicend}
\algtext*{bindent}
\algtext*{eindent}
\newcommand{\mathsc}[1]{{\normalfont\textsc{#1}}}
\definecolor{codegray}{rgb}{0.5,0.5,0.5}
\definecolor{backcolour}{rgb}{0.95,0.95,0.92}

\lstdefinestyle{mystyleNumber}{
    backgroundcolor=\color{backcolour},   
    numberstyle=\tiny\color{codegray},
    breakatwhitespace=false,         
    breaklines=true,                 
    captionpos=b,                    
    keepspaces=true,                 
    numbers=left,                    
    numbersep=5pt,                  
    showspaces=false,                
    showstringspaces=false,
    showtabs=false,                  
    tabsize=2
}

\lstdefinestyle{mystyleNoNumber}{
    backgroundcolor=\color{backcolour},   
    numberstyle=\tiny\color{codegray},
    breakatwhitespace=false,         
    breaklines=true,                 
    captionpos=b,                    
    keepspaces=true,                 
    showspaces=false,                
    showstringspaces=false,
    showtabs=false,                  
    tabsize=2
}

\lstdefinestyle{mystyleInLine}{
    breakatwhitespace=false,         
    breaklines=true,                 
    keepspaces=true,                 
    showspaces=false,                
    showstringspaces=false,
    showtabs=false,                  
    tabsize=2
}

\lstdefinelanguage{ASM}
{keywords={rule, if, then, endif, else, forall, choose, in, do, let, par, endpar, seqblock, endseqblock, derived, function, universe, where, or, and, not, with, seq, for, iterate, by, import},
sensitive=false,
mathescape=true
}

\begin{document}
\title{An Approach for Safe and Secure Software Protection Supported by Symbolic Execution\thanks{The research reported in this paper has been funded by BMK, BMDW, and the State of Upper Austria in the frame of the COMET Module Dependable Production Environments with Software Security (DEPS) within the COMET - Competence Centers for Excellent Technologies Programme managed by Austrian Research Promotion Agency FFG.
The final publication is available at Springer via \url{https://doi.org/10.1007/978-3-031-39689-2_7}}}
% The research reported in this {paper | article | chapter | monograph | master thesis | bachelor thesis | phd thesis | ... } has been [partly] funded by the Federal Ministry for Climate Action, Environment, Energy, Mobility, Innovation and Technology (BMK), the Federal Ministry for Digital and Economic Affairs (BMDW), and the State of Upper Austria in the frame of the SCCH competence center INTEGRATE [(FFG grant no. 892418)] in the COMET - Competence Centers for Excellent Technologies Programme managed by Austrian Research Promotion Agency FFG.
% The research reported in this {paper | article | master thesis | … } has been funded by the Federal Ministry for Climate Action, Environment, Energy, Mobility, Innovation and Technology (BMK), the Federal Ministry for Digital and Economic Affairs (BMDW), and the State of Upper Austria in the frame of the COMET Module Dependable Production Environments with Software Security (DEPS) within the COMET - Competence Centers for Excellent Technologies Programme managed by Austrian Research Promotion Agency FFG.

%
\titlerunning{An Approach for Safe and Secure Software Protection}
\author{Daniel Dorfmeister\inst{1}\orcidID{0000-0002-2718-6007} 
\and
Flavio Ferrarotti\inst{1}\orcidID{0000-0003-2278-8233} 
\and
Bernhard Fischer\inst{1}\orcidID{0000-0001-9737-0056}
\and
Evelyn Haslinger\inst{2} 
\and
Rudolf Ramler\inst{1}\orcidID{0000-0001-9903-6107}
\and 
Markus Zimmermann\inst{2}
}
\authorrunning{D. Dorfmeister et al.}
\institute{Software Competence Center Hagenberg, Austria\\ 
  \email{\{daniel.dorfmeister,flavio.ferrarotti,bernhard.fischer,rudolf.ramler\}@scch.at}
  \and
   Symflower GmbH, Linz, Austria\\
\email{\{evelyn.haslinger,markus.zimmermann\}@symflower.com}}
\maketitle              % typeset the header of the contribution
\begin{abstract} 
We introduce a novel copy-protection method for industrial control software. With our method, a program executes correctly only on its target hardware and behaves differently on other machines. The hardware-software binding is based on Physically Unclonable Functions (PUFs). 
We use symbolic execution to guarantee the preservation of safety properties if the software is executed on a different machine, or if there is a problem with the PUF response.
Moreover, we show that the protection method is also secure against reverse engineering.
%\keywords{First keyword  \and Second keyword \and Another keyword.}
\end{abstract}
\section{Introduction}
Industrial-scale reverse engineering is a serious problem, with estimated annual losses for industry at 6.4 billion euros in Germany alone\footnote{VDMA Product Piracy 2022 (\url{https://www.vdma.org/documents/34570/51629660/VDMA+Study+Product+Piracy+2022_final.pdf}). Last accessed: 30/01/2023.}.
Typically, the main effort needed to steal the intellectual property (IP) of companies producing machines with key software components resides in replicating the hardware.
Contrary, software can often be copied verbatim with no reverse engineering required.

In the DEPS\footnote{\url{https://deps.scch.at}} (short for Dependable Production Environments with Software Security) project we investigate new approaches to prevent the described IP theft.
In this paper, we explore binding a given program $P$ to a specific target machine $M$, so that $P$ only executes correctly on $M$.
If $P$ runs on a machine $M'$ other than $M$ (even if $M'$ is a replica of $M$), then $P$ should behave differently, but still meet the required safety constraint for industrial applications.
Our objective is to make the task of reverse engineering the protection extremely difficult and time consuming, rendering it uneconomical for the attacker.

%A PUF gives unique responses to specific challenges.
%The responses are determined by the inherent randomness introduced during manufacturing of different hardware components.
%PUFs act as a kind of fingerprint for the hardware since it cannot be replicated.

We make use of Physically Unclonable Functions (PUFs), a hardware-based security primitive~\cite{gassend2002silicon,herder2014physical,McGrath2019}.
Minor variations in the manufacturing process of a hardware component cause unintended physical characteristics.
Thus, this unique \emph{digital fingerprint} cannot be cloned easily.
A PUF uses these unique hardware characteristics to provide hardware-specific responses to user-defined challenges.
PUFs can use designated hardware, but they can also be based on standard components like Dynamic Random Access Memory (DRAM)~\cite{Keller2014,Sutar2016,Xiong2016,Kim2018}.
A common application of PUFs is for hardware-software binding~\cite{guajardo2007fpga,kumar2008butterfly,kohnhauser2015puf,xiong2019software}. 

In the environment targeted by DEPS, it is important that the alternative behavior of a protected program is always safe, in the sense that the program may run differently on different than the original hardware (making it difficult for an attacker to reverse engineer the protection), but it should do so without producing any harm.
Moreover, if a protected program running on the target machine receives an (unlikely but nevertheless possible) incorrect response to a given PUF challenge, the program must still behave safely.
Through the approach presented in this paper, we can attain this high degree of safety for a wide class of algorithms by applying symbolic execution techniques.
At the same time, we can also ensure a high level of security, in the sense that reverse engineering the protected software becomes an extremely difficult (and expensive) task, without guarantee of success.

Xiong et al.~\cite{xiong2019software} have previously presented a software protection mechanism in which they use dynamic PUFs (based on DRAM).
They bind software to hardware, protecting it against tampering by considering the timing of the software for the PUF response.
For the protection itself, they use self-checksumming code instances.
With the help of a PUF response, the checksum, and a reference value, they determine a jump address.
This can, however, result in unsafe behavior of the software or a crash, since the jump can be to a random function if the response of the PUF is not the expected one. 
We follow a different safe-by-design approach, where we only allow moves to next states that preserve safety constraints.

The verification of software safety by means of symbolic execution is not new---see, e.g.,~\cite{Ahmed2019}.
However, to the best of our knowledge, it has not been used to support safe copy-protection using software to hardware binding yet. 

The paper is organized as follows:
In \autoref{sec:prelim}, we briefly introduce the class of algorithms targeted by our protection method.
This class is captured by control state Abstract State Machines (ASMs), which is also a convenient formal specification method to precisely describe the proposed protection mechanism.
Note that ASMs can be considered as executable abstract programs and thus can also be symbolically executed. 
%This is followed by two short sections, namely Sections~\ref{sec:pufs} and~\ref{threatModel}, where briefly comment on different suitable PUFs and fix our threat model assumptions, respectively. 
We specify our threat model in \autoref{threatModel}.
Our main contribution is condensed in \autoref{sec:protection}, where we introduce the proposed protection mechanism through an example and then generalize it to turn any ASM specification of a control state algorithm into a copy-protected (safe) specification bound to a target hardware through a suitable PUF.
This is followed by a preliminary evaluation of the security of our protection method in \autoref{sec:security-evaluation}.
We conclude our presentation with a brief summary and future research plans in \autoref{summary}.

%\todo{Finish the introduction, (e.g., comment on why these type of Cyber physical systems are safety critical, describe briefly how the protection that we propose works, how we use symbolic execution to ensure safety, etc.) and add a paragraph regarding organization of the paper. Two pages maximum for the whole intro.}

\section{Control State ASMs}
\label{sec:prelim}

We use Abstract State Machines (ASMs) to formally specify the proposed copy-protection mechanism and to show that it is applicable to the whole class of algorithms captured by \emph{control state} ASMs~\cite{boerger:2003}.
This is a particularly frequent class of ASMs that represents a normal form for UML activity diagrams.
It allows the designer to define control Finite-State Machines (FSMs) with synchronous parallelism and the possibility of manipulating data structures.
Moreover, industrial control programs (i.e., our target in DEPS) belong to this class~\cite{BorgerR18}.
This paper can be understood correctly by reading ASM rules as pseudocode over abstract data types.
Next, we briefly review some of the basic ASM concepts. Standard reference books for ASMs are~\cite{boerger:2003,BorgerR18}.

An ASM of some signature $\Sigma$ can be defined as a finite set of transition rules of the form ${\bf if}$ \emph{Condition} $\bf then$ \emph{Updates}, which transforms states.
The condition or guard under which a rule is applied is a first-order logic sentence of signature $\Sigma$. 
\emph{Updates} is a finite set of assignments of the form $f(t_1, \ldots, t_n) := t_0$, which are executed in parallel.
The execution of $f(t_1, \ldots, t_n) := t_0$ in a given state $S$ proceeds as follows: 
(1) all parameters $t_0, t_1, \ldots, t_n$ are assigned their values, say $a_0, a_1, \ldots, a_n$,
(2) the value of $f(a_1, \ldots, a_n)$ is updated to $a_0$, which represents the value of $f(a_1, \ldots, a_n)$ in the next state.
Pairs of a function name $f$ (fixed by the signature) and optional arguments $(a_1, \ldots, a_n)$ of dynamic parameter values $a_i$, are called locations.
They are the ASM concept of memory units, which are abstract representations of memory addressing.
Location value pairs $(l, a)$ %where $l$ is a location and $a$ is a value, 
are called updates, the basic units of state change.
 
The notion of an ASM run \emph{or} computation is an instance of the classical notion of the computation of transition systems.
An ASM computation step in a given state simultaneously executes all updates of all transition rules whose guard is true in the state.
If and only if these updates are consistent, the result of their execution yields a next state.
A set of updates is consistent if it does not contain pairs $(l, a), (l, b)$ of updates to a common location $l$ with $a \neq b$.
Simultaneous execution, as obtained in one step through the execution of a set of updates, provides a useful instrument for high-level design to locally describe a global state change.
%This synchronous parallelism is further enhanced by the transition rule ${\bf forall}$ $x$ ${\bf with}$ $\varphi$ ${\bf do}$ $r$ which expresses the simultaneous execution of a rule $r$ for each $x$ satisfying a given condition $\varphi$. Similarly, 
Non-determinism, usually applied as a way of abstracting from details of scheduling of rule executions, can be expressed by the rule ${\bf choose}$ $x$ $\bf with$ $\varphi$ $\bf do$ $r$, which means that $r$ should be executed with an arbitrary $x$ chosen among those satisfying the property $\varphi$.
We sometimes use the convenient rule ${\bf let}$ $x = t$ $\bf in$ $r$, meaning: assign the value of $t$ to $x$ and then execute $r$. 

A \emph{control state} ASM (see Def.2.2.1. in~\cite{boerger:2003}) is an ASM whose rules are all of the form defined in \autoref{controlStateASM}.
Note that if there is no condition $\mathit{cond}_i$ satisfied for a given control state $i$, then these machines do not switch state.
There can only be a finite number of $\mathit{ctlState} \in \{1, \ldots, m\}$.
In essence, they act as internal states of an FSM and can be used to describe different system \emph{modes}.

% \newpage
\begin{lstlisting}[language=ASM, style=mystyleNumber, caption={General Form of Control State ASM Rules}, label = controlStateASM]
if $\mathit{ctlState} = i$ then 
    $\mathit{rule}$
    if $\mathit{cond}_1$ then
        $\mathit{ctlState} := j_1$
    $\cdots$
    if $\mathit{cond}_n$ then
        $\mathit{ctlState} := j_n$
\end{lstlisting}

Industrial control programs are usually required to satisfy certain safety and security properties.
The main challenge for the copy protection method proposed in this paper is to ensure that these properties are maintained. Our method achieves this using symbolic execution~\cite{King76,Pasareanu}.
Symbolic execution applies to ASMs the same way it does to high-level programming languages~\cite{PAUN201711251}.
This is not surprising as ASMs are executable abstract programs~\cite{abs-2209-06546}.

\emph{Symbolic execution} abstractly executes a program covering  multiple possible inputs of the program that share a particular execution path through the code.
In concrete terms, instead of considering normal runs in which all locations take actual values from the base set of the ASM states, we consider runs in which some state locations may take symbols (depicted by lowercase Greek letters) representing arbitrary values.
ASM runs will proceed as usual, except that the interpretation of some expressions will result in symbolic formulas.

\section{Threat Model}\label{threatModel}

Our threat model focuses on IP theft.
More precisely, on reverse engineering the obfuscated flow of a program.
In this work, we assume a kind of white-box manat-the-end scenario where an attacker is able to successfully gain access to an arbitrary number of protected programs.
The attacker does not have direct access to the source code but can analyze and decompile the binaries to obtain it.
They can also obtain the exact specification of the hardware on which the program runs and rebuild it.
The attacker's goal is to obtain information about the correct program flow and to rebuild or copy it.

This threat model and the related security of the protection mechanism presented in the following will be further discussed in \autoref{sec:security-evaluation}.
We also address what happens when the attacker gains access to a protected system, i.e., to the hardware the software is bound to.

\section{PUF-Based Software Protection Method}
\label{sec:protection}

This section introduces the key ideas and contribution of the paper. In \autoref{example}, we present our novel approach to copy-protect programs specified as control state ASMs through a simple example.
This illustrates our proposal to use PUFs to bind the correct execution of a control program to a target machine.
Moreover, it illustrates how symbolic execution can be used to ensure that the protected control ASM satisfies all safety constraints of the original control ASM. In \autoref{method}, we condense the approach into a precise method that can be automated.
Given an appropriate PUF, our method can turn any given control state ASM specification into an equivalent specification that will only run correctly on the target machine and satisfy the required safety constraints. 

For our approach to work, the following requirements must be met by the chosen PUF:
\begin{itemize}
    \item It must be possible to query the PUF at run-time and its response time should not adversely affect the functionality of the target program. 
    \item No error correction is required, but the response of the PUF must be reasonably reliable so that the software executes correctly on the target machine most of the time.
    Ultimately, this will depend on the fault-tolerance requirement of the specific application.
    \item The number of different challenge-response pairs of the PUF should ideally be greater than or equal to the number of control states of the program that one would like to protect with our method.
    A smaller number would also work, but it would make the protection less secure against reverse  engineering.
\end{itemize}

\subsection{Approach}\label{example}

\begin{figure}
    \centering
    \includegraphics[clip, trim=3cm 4.9cm 0cm 2cm, width=12cm]{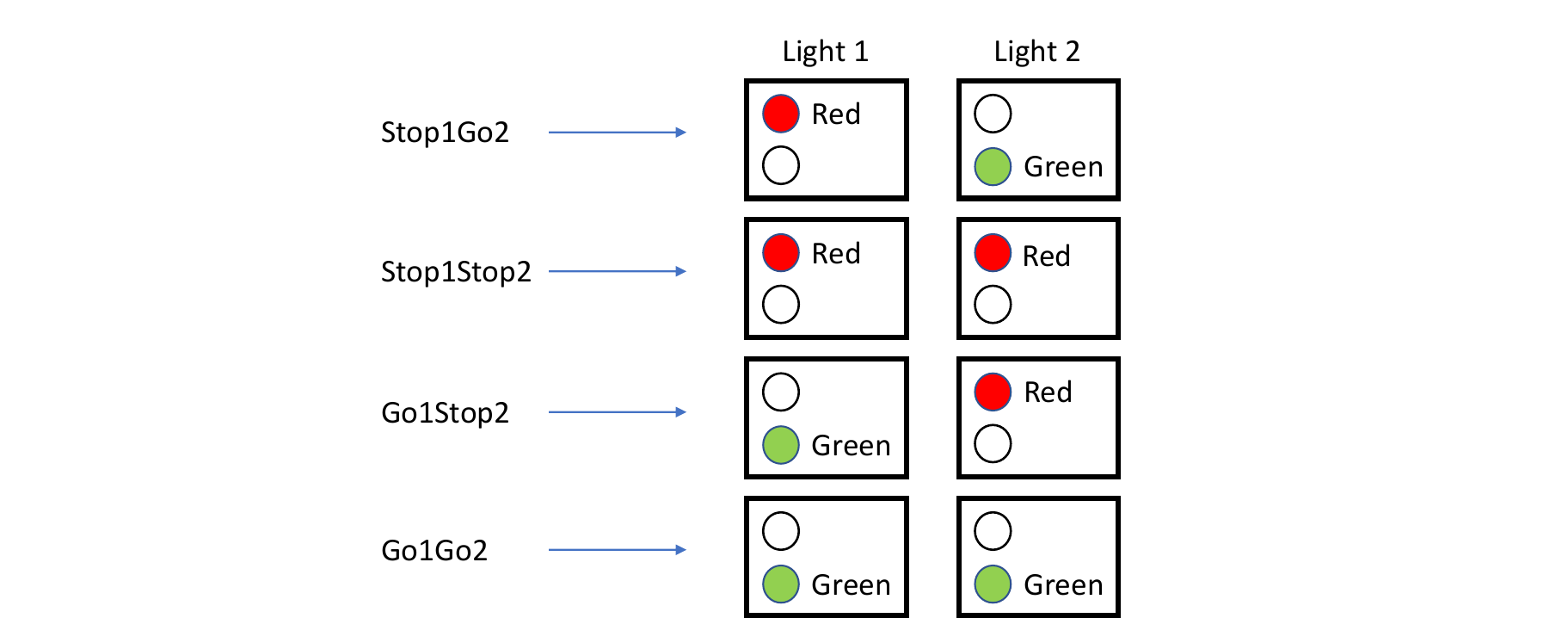}
    \caption{Possible states of one-way stop-go traffic lights}
    \label{fig:onewaytrafficlightstates}
\end{figure}

We introduce our approach through a simple example. Consider the ASM specification in~\cite{BorgerR18} (cf. \autoref{sec:prelim}) of a one-way traffic light control algorithm.
The proper behavior of this algorithm is defined by the ASM rule in \autoref{lst:1WayTrafficLight}.
For this example, there are $4$ possible control states as shown in \autoref{fig:onewaytrafficlightstates}.
Only the first three are used by the correct specification.
The fourth possible control state (i.e., $\mathit{Go1Go2}$) represents an unsafe, undesirable behavior.

\begin{lstlisting}[language=ASM, style=mystyleNumber, caption={One-Way Traffic Light: Correct Specification}, label = lst:1WayTrafficLight]
$\mathsc{1WayStopGoLight}=$
if $\mathit{phase} \in \{\mathit{Stop1Stop2}, \mathit{Go1Stop2}\}$ and $\mathit{Passed}(\mathit{phase})$ then	
	$\mathit{StopLight}(1) := \neg \mathit{StopLight}(1)$	
	$\mathit{GoLight}(1) := \neg \mathit{GoLight}(1)$	
	if $\mathit{phase} = \mathit{Stop1Stop2}$ then
		$\mathit{phase} := \mathit{Go1Stop2}$
	else
		$\mathit{phase} := \mathit{Stop2Stop1}$
if $\mathit{phase} \in \{\mathit{Stop2Stop1}, \mathit{Go2Stop1}\}$ and $\mathit{Passed}(\mathit{phase})$ then	
	$\mathit{StopLight}(2) := \neg \mathit{StopLight}(2)$	
	$\mathit{GoLight}(2) := \neg \mathit{GoLight}(2)$	
	if $\mathit{phase} = \mathit{Stop2Stop1}$ then
		$\mathit{phase} := \mathit{Go2Stop1}$
	else
		$\mathit{phase} := \mathit{Stop1Stop2}$ 
\end{lstlisting}

\newpage
We propose to bind this control state algorithm to the hardware using the sub-rule $\mathsc{ChoosePhase}$ in \autoref{lst:choosePhase}, which updates the value of $\mathit{phase}$ to the correct next phase shown in \autoref{lst:1WayTrafficLight} only if the algorithm is executed on the target hardware.
Otherwise, $\mathsc{ChoosePhase}$ updates $\mathit{phase}$ by choosing non-deterministically among the set of $\mathit{safePhases}$, i.e., the phases that will not lead to $\mathit{GoLight(1)} = \mathit{GoLight(2)} = \mathit{true}$, and thus violate the safety constraint.
% to the violation of the safety constraint for this example.
This procedure ensures that if the software is copied verbatim to a machine other than the target one, it will behave safely but also incorrectly and nondeterministically, making it challenging for an attacker to reverse engineer the protection.

\begin{lstlisting}[language=ASM, style=mystyleNumber, caption={Software-to-Hardware Binding Procedure}, label = lst:choosePhase]
$\mathsc{ChoosePhase}=$
let $\mathit{nextPhase} = \mathit{queryPUF}(\mathit{phase})$ in
	if $\mathit{nextPhase} \in \mathit{safePhases}$ then 
		$\mathit{phase} := \mathit{nextPhase}$
	else
		choose $x \in \mathit{safePhases}$ do 		
			$\mathit{phase} := x$
\end{lstlisting}

We use symbolic execution to systematically determine the set of $\mathit{safePhases}$, specifically (1) to determine the necessary conditions for a next state to be safe, and (2) to verify that this is indeed the case.
Note that we must ensure that the protection mechanism will not introduce any bug that leads to an unsafe state.
First, we assign a symbolic value to all relevant state locations:\vspace{-0.2cm}
\begin{align*}
L_0 = \{&\mathit{phase} \mapsto \alpha, \mathit{Passed}(\mathit{phase}) \mapsto \beta, \mathit{StopLight}(1) \mapsto \gamma, \mathit{GoLight}(1)  \mapsto \neg \gamma,\\
& \mathit{StopLight}(2) \mapsto \delta, \mathit{GoLight}(2) \mapsto \neg \delta\}
\end{align*}
Note that the relevant locations (or equivalently locations of interest) are all those that correspond to dynamic function names that appear in the conditional statements of $\mathsc{1WayStopGoLight}$. This might be in abundance since there could be locations that are not relevant to preserve the required safety constraint, i.e., that $\neg (\mathit{GoLight}(1) \wedge \mathit{GoLight}(2))$.
The problem is that if we only consider a sub-set of locations, there is the possibility that we miss some execution path that leads to a violation of the safety constraint that we want to preserve.
In general, this is undecidable and compromises might be necessary for complex applications.  

Next, we look at all value path conditions resulting from executing the \emph{protected} rule $\mathsc{1WayStopGoLight}$ (i.e., $\mathsc{Protected1WayStopGoLight}$) obtained by replacing lines $6, 8, 13$ and $15$ in $\mathsc{1WayStopGoLight}$ with the sub-rule $\mathsc{ChoosePhase}$.
W.l.o.g., in the symbolic execution we can simply assume that $\mathsc{ChoosePhase}$ assigns a symbolic value $\alpha'$ (possibly different from $\alpha$) to $\mathit{phase}$, since, at this point, we are only interested in determining the path conditions where the current value $\alpha$ of $\mathit{phase}$ can lead to $\mathit{GoLight(1)} = \mathit{GoLight(2)} = \mathit{true}$ in the next state.
This gives us $4$ path conditions and $4$ corresponding symbolic state locations, one for each of the $4$ possible symbolic step transitions $a, b, c$, and $d$ (executions) of $\mathsc{Protected1WayStopGoLight}$. 
\begin{itemize}

\item Path conditions $P_a$ and set of symbolic state locations $L_a$ of transition $a$:\vspace{-0.2cm} 
\begin{align*}
P_a \equiv &(\alpha = \mathit{Stop1Stop2} \vee \alpha = \mathit{Go1Stop2}) \wedge \beta \wedge  \alpha = \mathit{Stop1Stop2}\\
L_a  = &\{\mathit{phase} \mapsto \alpha', \mathit{Passed}(\mathit{phase}) \mapsto \beta, \mathit{StopLight}(1) \mapsto \neg\gamma, \mathit{GoLight}(1)  \mapsto \gamma,\\
& \mathit{StopLight}(2) \mapsto \delta, \mathit{GoLight}(2) \mapsto \neg \delta\}
\end{align*}

\item Path conditions $P_b$ and set of symbolic state locations $L_b$ of transition $b$: \vspace{-0.2cm}
\begin{align*}
P_b \equiv &(\alpha = \mathit{Stop1Stop2} \vee \alpha = \mathit{Go1Stop2}) \wedge \beta \wedge  \alpha \neq \mathit{Stop1Stop2}\\
L_b  = &\{\mathit{phase} \mapsto \alpha', \mathit{Passed}(\mathit{phase}) \mapsto \beta, \mathit{StopLight}(1) \mapsto \neg\gamma, \mathit{GoLight}(1)  \mapsto \gamma,\\
& \mathit{StopLight}(2) \mapsto \delta, \mathit{GoLight}(2) \mapsto \neg \delta\}
\end{align*}

\item Path conditions $P_c$ and set of symbolic state locations $L_c$ of transition $c$: \vspace{-0.2cm}
\begin{align*}
P_c \equiv &(\alpha = \mathit{Stop2Stop1} \vee \alpha = \mathit{Go2Stop1}) \wedge \beta \wedge  \alpha = \mathit{Stop2Stop1}\\
L_c  = &\{\mathit{phase} \mapsto \alpha', \mathit{Passed}(\mathit{phase}) \mapsto \beta, \mathit{StopLight}(1) \mapsto \gamma, \mathit{GoLight}(1)  \mapsto \neg \gamma,\\
& \mathit{StopLight}(2) \mapsto \neg \delta, \mathit{GoLight}(2) \mapsto \delta\}
\end{align*}

\item Path conditions $P_d$ and set of symbolic state locations $L_d$ of transition $c$: \vspace{-0.2cm}
\begin{align*}
P_d \equiv &(\alpha = \mathit{Stop2Stop1} \vee \alpha = \mathit{Go2Stop1}) \wedge \beta \wedge  \alpha \neq \mathit{Stop2Stop1}\\
L_d  = &\{\mathit{phase} \mapsto \alpha', \mathit{Passed}(\mathit{phase}) \mapsto \beta, \mathit{StopLight}(1) \mapsto \gamma, \mathit{GoLight}(1)  \mapsto \neg \gamma,\\
& \mathit{StopLight}(2) \mapsto \neg \delta, \mathit{GoLight}(2) \mapsto \delta\}
\end{align*}
\end{itemize}

In the next step, we group the path conditions that lead to the same symbolic state. In our running example, we see that whenever the condition $P_a \vee P_b$ is satisfied, we get the symbolic state $L_a = L_b$. Likewise, whenever the condition $P_c \vee P_d$ is satisfied, we get the symbolic state $L_c = L_d$. Using standard logical equivalences, we can simplify the path conditions as follows:\vspace{-0.2cm} 
\begin{align*}
P_a \vee P_b &\equiv (\alpha = \mathit{Stop1Stop2} \vee \alpha = \mathit{Go1Stop2}) \wedge \beta\\
P_c \vee P_d &\equiv (\alpha = \mathit{Stop2Stop1} \vee \alpha = \mathit{Go2Stop1}) \wedge \beta
\end{align*}
Moreover, if we restrict our attention to the locations of interest and their symbolic values following a step transition, we can clearly see that in the next state the assertion $\mathit{GoLight(1)} = \mathit{GoLight(2)} = \mathit{true}$ holds, i.e., the safety constraint is violated, iff $\mathit{Passed}(\mathit{phase})$ and
%we get that the 
%the following (expressed in the logic for ASMs):
\vspace{-0.2cm}
\begin{align*}
&\big((\mathit{phase} = \mathit{Stop1Stop2} \vee \mathit{phase} = \mathit{Go1Stop2}) \wedge \neg GoLight(1) \wedge GoLight(2)\big) \vee\\
&\big((\mathit{phase} = \mathit{Stop2Stop1} \vee \mathit{phase} = \mathit{Go2Stop1}) \wedge \neg GoLight(2) \wedge GoLight(1)\big) 
\end{align*}
%\begin{align*}
%\exists X \big(&\mathit{upd}_R(X) %\wedge [X](\mathit{GoLight(1)} \wedge %\mathit{GoLight(2)}\big) %\leftrightarrow \big(\mathit{Passed}%(\mathit{phase}) \wedge\\
%&\big((\mathit{phase} = %\mathit{Stop1Stop2} \vee %\mathit{phase} = \mathit{Go1Stop2}) %\wedge \neg GoLight(1) \wedge %GoLight(2)\big) \vee\\
%&\big((\mathit{phase} = %\mathit{Stop2Stop1} \vee %\mathit{phase} = \mathit{Go2Stop1}) %\wedge \neg GoLight(2) \wedge %GoLight(1)\big)\big) 
%\end{align*}
Finally, we get that $\mathit{safePhases}$ is the set  \[\{x \in \{\mathit{Stop1Stop2}, \mathit{Go1Stop2}, \mathit{Stop2Stop1}, \mathit{Go2Stop1}\} \mid \neg \mathrm{Cond}(x)\}\]
where 
\begin{align*}
\mathrm{Cond}(x) \equiv &\big((x = \mathit{Stop1Stop2} \vee x = \mathit{Go1Stop2}) \wedge \neg GoLight(1) \wedge GoLight(2)\big) \vee\\
&\big((x = \mathit{Stop2Stop1} \vee x = \mathit{Go2Stop1}) \wedge \neg GoLight(2) \wedge GoLight(1)\big)\big)
\end{align*}

\subsection{Method}\label{method}

We now generalize the example from the previous subsection to a precise method that can be automated. Let $P$ be a control state ASM (i.e., an executable abstract program) with finitely many control states $ctlState \in \{1, \ldots, m\}$, whose rules $r_1, \ldots, r_m$ are all of the form defined in \autoref{controlStateASM}. We assume w.l.o.g. that each $r_i$ is guarded by ${\bf if} \, \mathit{ctlState} = i \, {\bf then} \, r_i$. We define:\vspace{-0.2cm} 
\[A = \{(i,j) \mid i,j \in \{1, \ldots, m\} \wedge  \mathit{subRuleOf}(\mathit{ctlState} := j, r_i)\}\]
where $\mathit{subRuleOf}(s, r)$ is a Boolean function that evaluates to true only if $s$ is a sub-rule of $r$. That is, $(i,j) \in A$ iff the rule $r_i$ updates the control state to $j$. 

The mechanical steps to copy-protect $P$ with safety property $\Psi$ using a (suitable) PUF $\mathit{puf}: C \rightarrow R$, with the set of challenges $C$ and corresponding responses $R$, are described in the following $7$ steps.
Note that Step~1 is simply a prerequisite to implement the function $\mathit{queryPUF}$ used in \autoref{lst:choosePhase} in the one-way traffic lights example introduced in the previous section.
Indeed, the $\mathsc{ChoosePhase}$ rule results from applying Step~2 of our method.
The next two steps in our method simply refine the application of the proposed protection based on updating control states via the PUF function, i.e., it corresponds to using $\mathsc{ChoosePhase}$ to select the next control state in our previous example.
Steps~5--7 are a formalization and generalization of the symbolic execution analysis done in the previous section to determine the safe control state transitions, that is to determine the set of $\mathit{safePhases}$ in our example. 

\begin{enumerate}
    \item Fix an injection $f: C' \rightarrow R$ such that $C' \subseteq C$, $|C'| = |A|$ and $f(x) = \mathit{puf}(x)$, and a corresponding bijection $g: A \rightarrow C'$.
    Note that this function $f$ is simply an injective restriction of $\mathit{puf}$ to a subdomain $C'$ of size $|A|$, as required to encode all possible control state transitions of $P$ (using the function $g$).
    If it is not possible to fix such an injective function $f$, this means that the function $\mathit{puf}$ is not ``big enough'' to encode all control state transitions.
    In this latter case, we simply consider that $\mathit{puf}$ is not suitable for our protection strategy.
    An alternative would be to apply the protection to a subset of control states, but this would unnecessarily complicate the presentation of the general method. 
    \item For each tuple $(x, y) \in A$, define $\mathsc{ChooseCtlState}(x,y)$ as in \autoref{lst:chooseCtlState}.
    
\begin{lstlisting}[language=ASM, style=mystyleNumber, caption={Choose next control state using the PUF}, label = lst:chooseCtlState]
$\mathsc{ChooseCtlState}(x,y)=$
let $\mathit{nextCtlState} = \mathit{puf}(g(x,y))$ in
	if $\mathit{nextCtlState} \in \mathit{safeCtlStates}$ then
		$\mathit{ctlState} := \mathit{nextCtlState}$
	else
		choose $z \in \mathit{safeCtrlStates}$ do
			$\mathit{ctlState} := z$
\end{lstlisting}

    \item For each control state rule $r_i$ ($i = 1, \ldots, m$) of $P$, replace every occurrence of a sub-rule of the form $\mathit{ctlState} := j$ by the $\mathsc{ChooseCtlState}(\mathit{ctrlState}, j)$ sub-rule defined in the previous step. 
    
    \item In each control state rule $r_i$ ($i = 1, \ldots, m$) of $P$, replace the condition $\mathit{ctlState} = i$ in its guard by $\mathit{ctlState} \in \{x \mid   \exists (y,z) \in A \,(g(y,z) = x \wedge y = i) \}$.

    \item Assign a different symbolic value to all locations in the initial state that correspond to terms that appear in some conditional statements of a ${\bf if}$-rule in $P$, i.e., to all locations of interest. This results in an assignment $\mathit{symVal}$, such that $\mathit{symVal}(h(\bar{t})) = \alpha_i$ iff $\alpha_i$ is the symbolic value assigned to the location $(h, \bar{t})$.
    
    \item Symbolically execute $P$ on the (symbolic) initial state built in the previous step. This results in a set $\mathit{Sym}$ of pairs of the form $(\varphi_i, L_i)$, where $\varphi_i$ is a path condition (as in standard symbolic execution) and $L_i$ is a set that maps each term (corresponding to a location with a symbolic value) to its symbolic value in the successor state of the initial one.

    \item Finally, set $\mathit{safeCtlStates} = \{x \in \{1, \ldots, m\} \mid \neg \mathrm{Cond}(x) \}$, 
    where \vspace{-0.2cm} \[\mathrm{Cond}(x) \equiv \bigvee_{(\varphi_i,L_i) \in \mathit{Sym}} \big(\varphi_i' \wedge \Psi(L_i)\big)\] and $\varphi_i'$ is the expression obtained by replacing the symbolic values in $\varphi_i$ by the terms corresponding to the locations with these symbolic values in the initial state.
    In turn, $\Psi(L_i)$ is obtained by first replacing in $\Psi$ (i.e., in the safety constraint) every term $t_j$ by the symbolic expression $L_i(t_j)$ and then replacing the symbolic values in the resulting expression by terms corresponding to the locations with these symbolic values in the initial state.  
\end{enumerate}

\section{Security Evaluation}
\label{sec:security-evaluation}

% I would say around 1 page? (How an attacker could reverse engineer this \rightarrow static, dynamic)

The safety of the proposed method for copy-protection of control state software is ensured by construction.
We now analyze its security. That is, assuming the threat model in \autoref{threatModel}, we evaluate how an attacker could circumvent the copy-protection.
To eliminate the protection, an attacker needs to perform an analysis that enables them to determine the correct value of $\mathit{nextCtlState}$ every time the rule in \autoref{lst:chooseCtlState} is called, or equivalently, whenever the current $\mathit{ctlState}$ is updated.
The attacker can only access a binary of the target program. 
%Now that the safety of the program, even in the case of incorrect behavior, has been ensured, it is still necessary to evaluate the security of the protection mechanism proposed.
%This is based purely on the idea and not yet on any concrete implementation.
%An attacker therefore has two possibilities: (i) analyzing the protected program statically and (ii) doing so dynamically. We will discuss both options hereinafter.
Thus, they will need to perform some kind of analysis, either of the binary, or of some decompiled version of it. The analysis can either be static or dynamic. Next, we discuss both possibilities. 

\noindent
\textbf{Static Analysis.}
We assume that an attacker can gain access to any number of protected binaries, compiled using different PUFs, i.e., for different systems.
This clearly does not provide any additional information regarding the values that $\mathit{nextCtlState}$ should take at different execution stages.
Indeed, the attacker cannot determine the value of the function $\mathit{puf}$ (see line~2 of \autoref{lst:chooseCtlState}) for a given challenge with static analysis. 
Even if an attacker can decompile the binary, which is not a trivial task for machine code, and then bypasses any obfuscation applied to it and gets a human readable version of the protected program, this still will not provide any information regarding the challenge/response values of $\mathit{puf}$. 
An alternative is for the attacker to understand the logic of the control state program through inspection of a decompiled high-level code. Then they could possibly determine the correct $\mathit{nextCtlState}$ for each of the control rules.
This task is time consuming and complex, even for the simple one-way traffic light control algorithm in \autoref{example}. In fact, writing the algorithm from scratch would probably be faster.
This could easily become insurmountable as the complexity of the control state program increases and consequently its logic and number of states.   
%the one-way traffic light control algorithm as well as the algorithm responsible for selecting the next phase.
%They know that the next phase is determined by a PUF response, and about the interface used for this purpose.
%However, the attacker has no information about the particular PUF responses.
%Thus, they can only guess which of the phases the algorithm chooses next.
%The additional binaries do not help the attacker either, since they possibly only contain other PUF challenges.
%The small example show in \autoref{sec:protection}, with only four possible phases, all potential program runs can be quickly probed, and assuming the expected behavior of the software, the correct program flow can be determined.
%However, this becomes ever more difficult with more complex programs, and is in consequence an exponential problem.
We can conclude that pure static analysis is not a real threat by itself for our protection method.

\noindent
\textbf{Dynamic Analysis.}
We consider two scenarios for dynamic analysis.
In the first one, the attacker has access to a system with identical specifications as the system the protected software is bound to.
This does not give any advantage to the attacker over the static analysis, since the PUF cannot be cloned.
Moreover, the runs of the program on this hardware will be non-deterministic w.r.t. $\mathit{nextCtlState}$.

In the second scenario, the attacker can run the program on the intended system, i.e., the unique system where the function $\mathit{puf}$ will return the correct responses.
Then, the attacker can trace the correct program flow via dynamic analysis.
This can still be time consuming, in particular if the system is complex with many possible execution paths.
Moreover, this dynamic analysis can interfere with PUF responses.
For example, operations on DRAM are time-dependent, including the Rowhammer PUF~\cite{SchallerXASG0S17,Anagnostopoulos18a}, for instance, where memory must be accessed at high frequency, which would be hindered by other processes.
%, since for this to work correctly memory must be accessed at high frequency as unhindered as possible by other processes.
The attacker could bypass this by querying the PUF separately from the protected program and then integrating the responses into the binary, but this would, once again, require a complex, time-consuming, and detailed analysis.
This relative weakness can be dealt with by using complementary obfuscation techniques and/or physical protection mechanisms.  

\section{Summary}\label{summary}

We introduced a novel method to bind control state software to specific hardware.
The method ties  the logic of the control state program to the unique responses provided by the PUF of the target hardware, so that it will only behave correctly if it is executed on the correct machine.
Otherwise, the program will behave differently and in a non-deterministic manner.
At the same time it will not crash, turning reverse engineering of the protection into a difficult, complex, and expensive task.
Moreover, our copy-protection method ensures by construction that the safety properties of the software are preserved, even when it is illegally copied and executed on a cloned machine.
This high level of safety is enabled by applying symbolic execution techniques and for a wide class of algorithms, namely for any algorithm that can be correctly specified by a control state ASM.

In future work, we plan to apply this protection method to an industrial case study, developing the necessary tools to automate the required tasks.
This will allow us to evaluate aspects such as scalability of the proposed method and concrete effort required to reverse engineer it.
It should be noted that step~7 of the method might result on long formulae whose evaluation could unacceptably affect the response time of a complex control state algorithm.
In principle, this can be dealt with by simplifying the resulting expressions using logical equivalences (as in the example presented in \autoref{example}).
Theorem provers in general, and SMT solvers in particular, can assist in this task.
The latter have a long history of successful application in conjunction with symbolic execution.

In its current initial form, our method for software protection requires to manually identify the control states of the target programs as well as a formal specification of the safety constraints that need to be preserved. Processes to automate the identification of control states that are suitable to apply the described protection are currently being investigated within the DEPS project. Safety constraints necessarily need to be translated into a symbolic form so that symbolic execution can be applied to prove their preservation. 

%\todo{Here we can describe briefly what we have shown in the paper and the next steps, including planned evaluation of scalability of the method (symbolic execution + evaluation of safe states)}
% Daniel: isn't relevant anymore now that we don't describe the µGlue PUF design
% repeatability of bit flips (noticing that given the safety assurance of our method, a high repeatability would be enough), functionality and performance)


\bibliographystyle{splncs04}
\bibliography{bibl.bib}

\end{document}